\documentclass{PoS}
\usepackage{amsmath}
\usepackage{bm}

\title{Electron-photon deep inelastic scattering at small $\bm{x}$ in holographic QCD}

\ShortTitle{Electron-photon DIS at small $x$ in holographic QCD}

\author{\speaker{Akira Watanabe}\\
        Institute of High Energy Physics and Theoretical Physics Center for Science Facilities, Chinese Academy of Sciences, Beijing 100049, People's Republic of China\\
        and\\
        University of Chinese Academy of Sciences, Beijing 100049, People's Republic of China\\
        E-mail: \email{akira@ihep.ac.cn}}

\author{Hsiang-nan Li\\
        Institute of Physics, Academia Sinica, Taipei 11529, Taiwan, Republic of China\\
        E-mail: \email{hnli@phys.sinica.edu.tw}}

\abstract{
We study the electron-photon deep inelastic scattering at small 
Bjorken variable $x$ in the framework of holographic QCD, 
employing the Pomeron exchange to describe the involved strong interaction.
With the Brower-Polchinski-Strassler-Tan Pomeron exchange 
kernel and appropriate wave functions for the incident and target 
particles, which are defined in the five-dimensional AdS space, 
we obtain the photon structure functions. It is shown that
our predictions agree with the experimental data measured 
at LEP and with those derived from a known parameterization of the 
photon parton distribution functions.
}

\FullConference{XXVI International Workshop on Deep-Inelastic Scattering and Related Subjects (DIS2018)\\
		16-20 April 2018\\
		Kobe, Japan}

\begin{document}

%%%%%%%%%%%%%%%%%%%%%%%%%%%%%%%%%%%
\section{Introduction}
%%%%%%%%%%%%%%%%%%%%%%%%%%%%%%%%%%%
Although a photon is a fundamental particle, it is possible to 
investigate its internal structure since a photon may fluctuate 
into quark-antiquark pairs in high energy scattering processes.
This investigation can be realized with the electron-photon deep
inelastic scattering (DIS), and the quark-gluon structure of 
the photon is explored through the structure functions.
Comprehensive reviews for photon structure functions were given in 
Refs.~\cite{Nisius:1999cv,Krawczyk:2000mf}.
Historically, the electron-photon DIS is a cleaner process 
compared to other high energy ones involving hadrons as the 
initial states, and the corresponding experimental data have provided theorists 
valuable opportunities to test the perturbative techniques of 
QCD in the region of a large Bjorken variable $x$.
The situation is completely different in the 
small $x$ region, where a photon can no longer be regarded as a point 
particle: its nature as a vector meson becomes dominant, which 
should be described by effective models, such as the vector meson dominance model.

In this work, we will study the 
photon structure functions at small $x$ in the framework of holographic QCD, which 
is an effective approach of QCD based on the AdS/CFT correspondence.
It is known that the Pomeron, corresponding to the closed string sector in the 
string theory, and identified as a graviton in the higher dimensional 
curved space, gives a dominant contribution in the small $x$ region. 
There are various descriptions of the Pomeron 
in the literature on holographic QCD. Here we adopt the Pomeron exchange kernel
proposed by Brower, Polchinski, Strassler, and Tan (BPST)~\cite{Brower:2006ea}.
Experimental data of the electron-nucleon DIS indicate that the Pomeron nature 
transits, depending on the probe photon four-momentum squared $Q^2$, from 
the so-called soft to hard Pomeron. It has been 
demonstrated~\cite{Brower:2010wf,Watanabe:2012uc,Watanabe:2013spa} that the 
BPST kernel, as applied to DIS at small $x$, well reproduces this transition.

A key issue is how to treat the target photon in the higher 
dimensional background, for which we take the following two approaches.
The wave function of the five-dimensional U(1) 
vector field~\cite{Polchinski:2002jw}, which couples to a lepton at the UV 
boundary, is applicable to the probe virtual photon. In our first approach, 
we utilize this U(1) vector 
field to describe the target photon with very tiny four-momentum 
squared~\cite{Watanabe:2015mia}.
The second approach is based on the vector meson dominance model:
we calculate the real photon structure functions by using the gravitational 
from factor of the $\rho$ meson constructed from a bottom-up 
AdS/QCD model~\cite{Abidin:2008ku}.
The three adjustable parameters of the model all reside in the BPST kernel.
Two of them, controlling the energy dependence of a cross section and 
the strength of the confinement effect, have been determined in the previous 
study on the nucleon DIS with the input of the proton  
structure function $F_2$ measured by HERA~\cite{Aaron:2009aa}.
Therefore, the results presented here can be regarded as our predictions, 
although the overall factor needs to be tuned to fit the experimental data 
of the OPAL collaboration at LEP~\cite{Abbiendi:2000cw}.

We show that our predictions in the two approaches are consistent with 
each other, and agree with the data and those derived from a known 
parameterization of the photon parton distribution functions (PDFs)~\cite{Gluck:1999ub}.
The above observation implies that the vector meson dominance is realized 
in the present model setup, and that our framework is appropriate for the study
of small $x$ physics. Besides, our predictions can be 
tested at future linear colliders, such as the planned International Linear Collider.

%%%%%%%%%%%%%%%%%%%%%%%%%%%%%%%%%%%
\section{Model setup}
\begin{figure}[tb]
\centering
\includegraphics[width=0.51\textwidth]{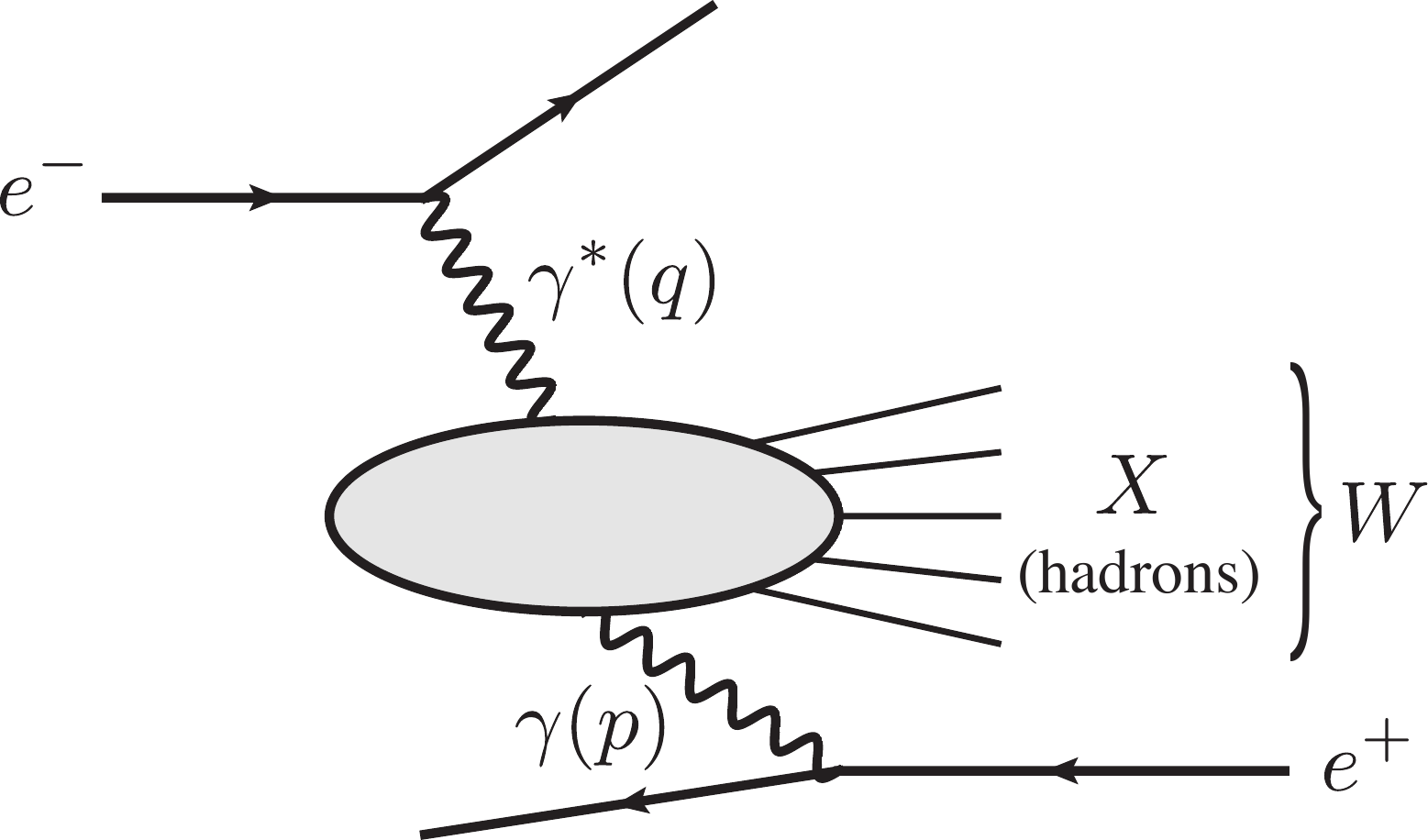}
\caption{Electron-photon deep inelastic scattering.}
\label{kinematics}
\end{figure}

%%%%%%%%%%%%%%%%%%%%%%%%%%%%%%%%%%%
We analyze the electron-photon DIS schematically shown in Fig.~\ref{kinematics},
where $q$ and $p$ denote the four-momenta of the probe and target photons, 
respectively, and $W$ is the invariant mass of the hadronic final state $X$.
For the real photon structure in the unpolarized case, we have
the differential cross section 
\begin{equation}
\frac{{{d^2}{\sigma_{e \gamma \to eX}}}}{{dxd{Q^2}}} = \frac{{2 \pi {\alpha^2}}}{{x{Q^4}}}\left[ {\left\{ {1 + {{\left( {1 - y} \right)}^2}} \right\}F_2^\gamma \left( {x,{Q^2}} \right) - {y^2}F_L^\gamma \left( {x,{Q^2}} \right)} \right], \label{DCS}
\end{equation}
in which $\alpha$ and $y$ are the fine structure constant and the inelasticity, respectively.
The Bjorken scaling variable $x$ is defined, in terms of $Q^2 = -q^2$ and $P^2 = -p^2$, by
\begin{equation}
x = \frac{Q^2}{ Q^2 + W^2 + P^2 }.
\end{equation}
Since we concentrate on the kinematic region with $W^2 \gg Q^2 \gg P^2$, 
the above definition can eventually be approximated by $x \approx Q^2/W^2$.

The two structure functions are expressed, with the 
BPST Pomeron exchange kernel $\chi$ in our model setup, as
\begin{equation}
F^{\gamma}_{i} (x,Q^2) = \frac{\alpha g_0^2 \rho^{3/2}Q^2}{32 \pi ^{5/2}}
\int dzdz' P_{13}^{(i)} (z,Q^2) P_{24}(z',P^2) (zz') \mbox{Im} [\chi (W^2,z,z')],\;\;i = 2,L\,, \label{SFs}
\end{equation}
where the overall factor $g_0^2$ controls the magnitude of the structure functions, 
and $\rho$ governs their energy dependence.
We employ the modified kernel for Eq.~\eqref{SFs}, 
\begin{align}
&\mbox{Im} [\chi_{mod} (W^2,z,z')] \equiv \mbox{Im} [\chi_c (W^2,z,z') ] + \mathcal{F} (z,z',\tau) \mbox{Im} [\chi_c(W^2,z,z_0^2/z') ],\label{MK} \\
&\mathcal{F} (z,z',\tau) = 1 - 2 \sqrt{\rho \pi \tau} e^{\eta^2} \mbox{erfc}( \eta ), \\
&\eta = \left( -\log \frac{zz'}{z_0^2} + \rho \tau \right) / {\sqrt{\rho \tau}}.
\end{align}
The first term on the right-hand side of Eq.~(\ref{MK}) is the kernel from 
the conformal field theory, 
\begin{equation}
\mbox{Im} [\chi_{c} (W^2,z,z') ] = e^{(1-\rho)\tau} e^ {-[({\log ^2 z/z'})/{\rho \tau}]} / \sqrt{\tau}, \label{CK}
\end{equation}
with $\tau = \log (\rho z z' W^2 /2)$. The second term on the right-hand side
of Eq.~(\ref{MK}) mimics the strength of the confinement effect adjusted by
the parameter $z_0$. 
It has been found~\cite{Watanabe:2012uc,Watanabe:2013spa} that the
results from Eq.~(\ref{MK}) are in better agreement with the data of the 
nucleon structure functions, compared to those from Eq.~(\ref{CK}).

The functions $P_{13}$ and $P_{24}$ in Eq.~\eqref{SFs} represent the density 
distributions of the colliding particles in the five-dimensional AdS space.
$P_{13}$ for the probe photon are set to the wave functions of the 
five-dimensional U(1) vector field, 
\begin{align}
P_{13}^{(2)} (z,Q^2) &= Q^2 z \left\{ K_0^2 (Qz) + K_1^2 (Qz) \right\}, \label{P13_2} \\
P_{13}^{(L)} (z,Q^2) &= Q^2 z K_0^2 (Qz). \label{P13_L}
\end{align}
As mentioned in the Introduction, 
we take the same function as Eq.~\eqref{P13_2} for the target photon 
wave functions $P_{24}$ with tiny four-momentum squared~\cite{Watanabe:2015mia} 
in the first approach. As to the vector meson dominance model
in the second approach, we consider the photon-$\rho$ meson scattering
with $P_{24}$ being described by the $\rho$ meson gravitational form factor 
in Ref.~\cite{Abidin:2008ku}.

%%%%%%%%%%%%%%%%%%%%%%%%%%%%%%%%%%%
\section{Numerical results}
\begin{figure}[tb]
\centering
\includegraphics[width=0.87\textwidth]{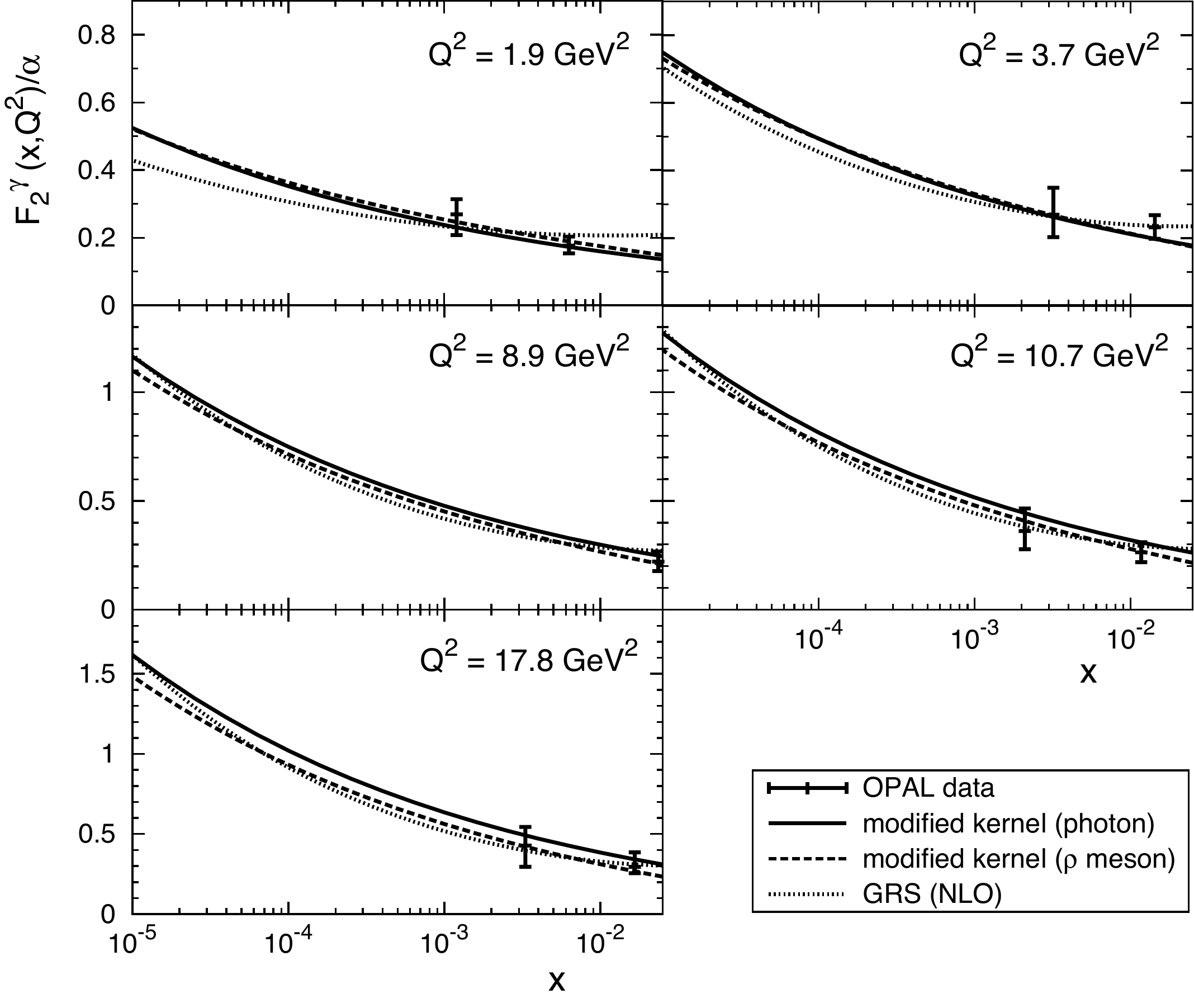}
\caption{
$F_2^\gamma (x,Q^2)$ as a function of the Bjorken variable $x$ for various $Q^2$.
In each panel, experimental data measured by the OPAL collaboration 
at LEP~\cite{Abbiendi:2000cw} are displayed with error bars,
the solid and dashed curves are from our calculations, and
the dotted curves are from the PDF parameterization~\cite{Gluck:1999ub}.
}
\label{photon_F2}
\end{figure}

%%%%%%%%%%%%%%%%%%%%%%%%%%%%%%%%%%%
We present in Fig.~\ref{photon_F2}
the behavior of the resulting photon structure function $F_2$ at small $x$.
The solid and dashed curves represent our predictions 
for the photon-photon and photon-$\rho$ meson scattering, respectively, 
which match the LEP data~\cite{Abbiendi:2000cw} within 
the errors, and with those derived from the GRS PDF set at 
next-to-leading-order accuracy~\cite{Gluck:1999ub}.
The agreement implies that the BPST Pomeron exchange kernel works 
well to describe the photon structure in the considered nonperturbative 
kinematic region, and that the vector meson dominance is realized in 
the present model setup.

\begin{figure}[tb]
\centering
\includegraphics[width=0.99\textwidth]{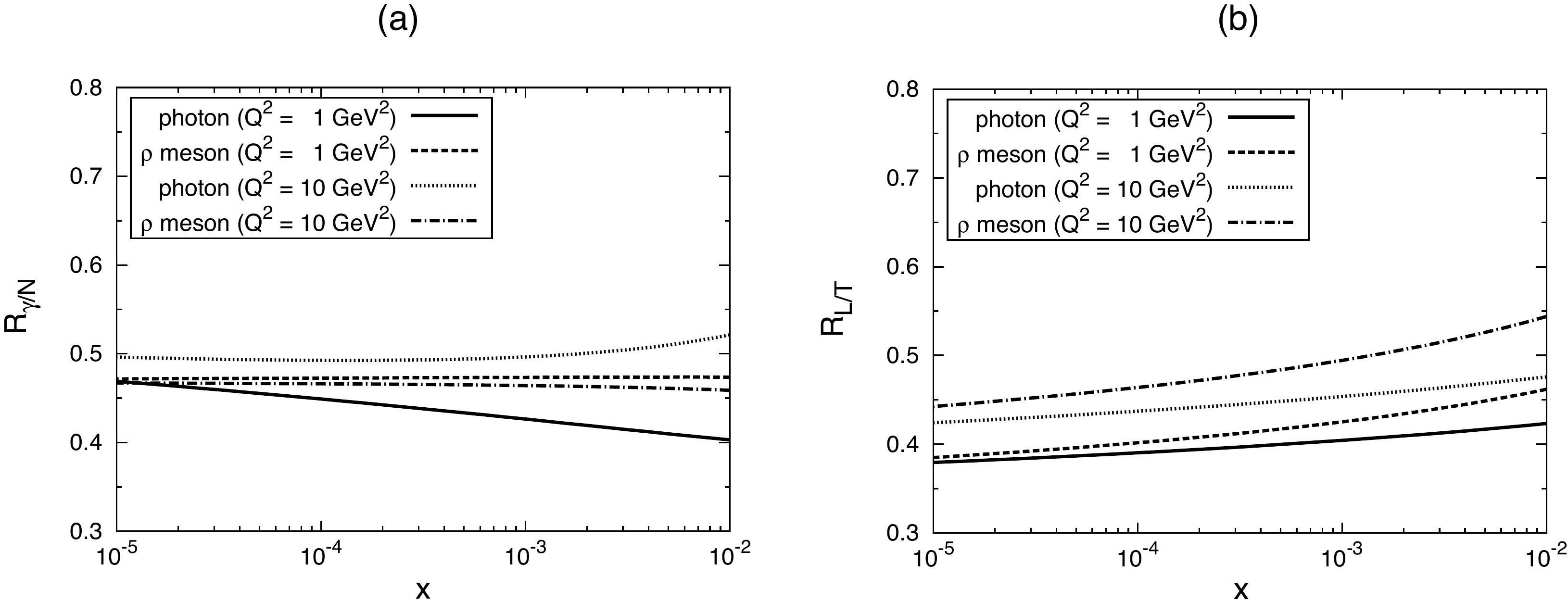}
\caption{
Ratios (a) $R_{\gamma / N} = F_2^\gamma (x,Q^2) / [\alpha F_2^N (x,Q^2)]$ 
and (b) $R_{L/T} = F_L (x,Q^2) / F_T (x,Q^2)$ as functions of 
the Bjorken variable $x$ for $Q^2 = 1$ and 10~GeV$^2$.
The nucleon results $F_2^N (x,Q^2)$ are taken from 
Ref.~\cite{Watanabe:2013spa}, and common to all the curves in the panel (a).
}
\label{SF_ratios}
\end{figure}

Some ratios of the structure functions are exhibited in Fig.~\ref{SF_ratios}
for a more detailed comparison between the two approaches.
The dependence of the ratio 
$R_{\gamma / N} = F_2^\gamma (x,Q^2) / [\alpha F_2^N (x,Q^2)]$ 
on the Bjorken variable $x$ is plotted in the panel (a), where the results for
the nucleon structure function $F_2^N (x,Q^2)$ are taken from 
Ref.~\cite{Watanabe:2013spa}. The ratios for the photon-photon scattering 
reveal opposite $x$ dependencies between the $Q^2 = 1$ and 10~GeV$^2$ 
cases, such that their difference increases with $x$.
Both the $x$ and $Q^2$ dependencies of the ratios for the 
photon-$\rho$ meson scattering are quite weak.
These features may be attributed to the fact that 
the $\rho$ meson is a normalizable mode, while the photon, described 
by Eq.~\eqref{P13_2}, is not. Note that the peak positions of the photon and 
$\rho$ meson density distributions are located in the UV (small $z$) 
and IR (large $z$) regions in the AdS space, respectively.
Since the larger $Q^2$ probe is sensitive to the distributions
at the smaller $z$, it is easy to understand that the ratio for the 
photon-photon scattering at $Q^2=10$~GeV$^2$ is higher than the others in 
the whole considered $x$ range.

\begin{figure}[tb]
\centering
\includegraphics[width=0.49\textwidth]{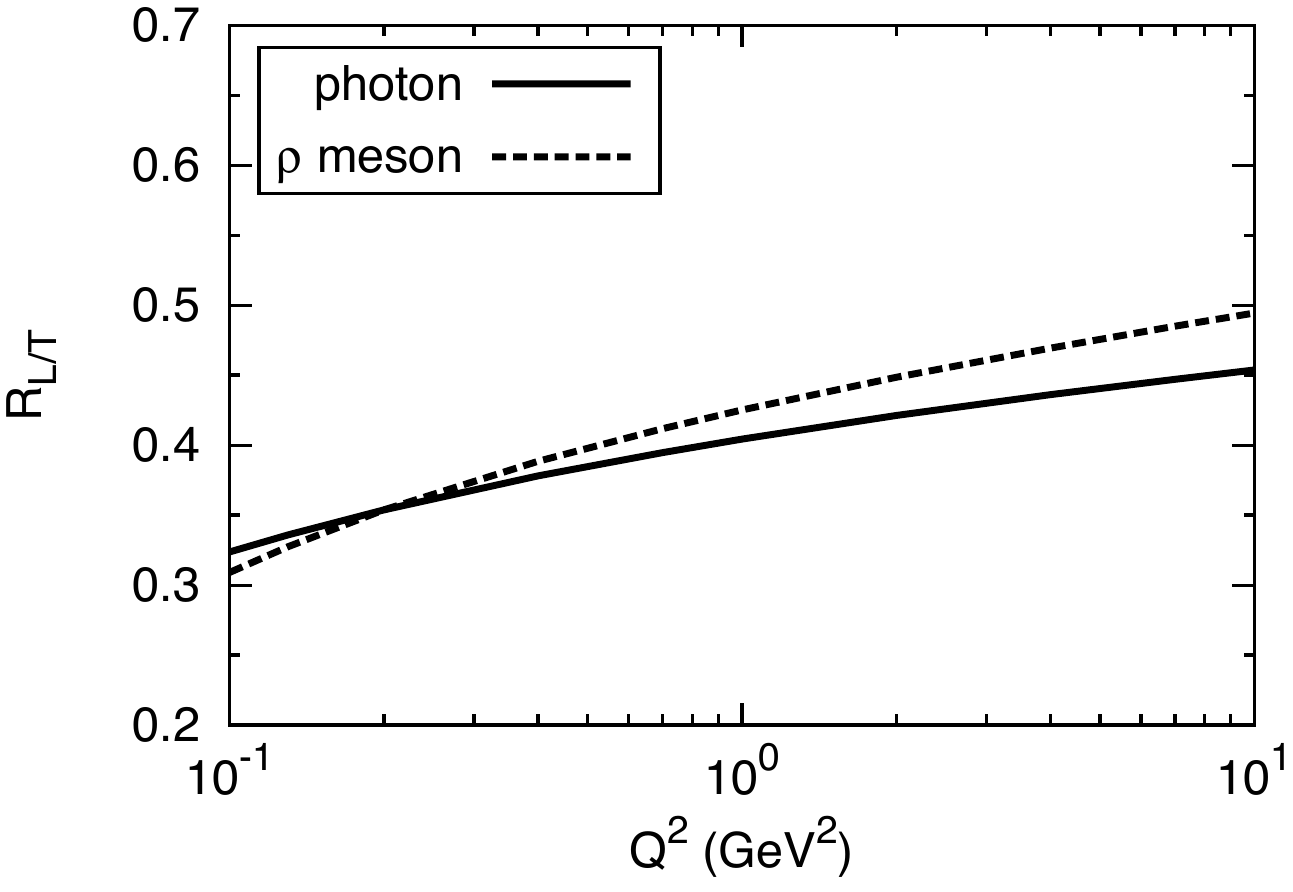}
\caption{
$R_{L/T} = F_L (x,Q^2) / F_T (x,Q^2)$ as a function of $Q^2$ at $x=10^{-3}$.
}
\label{L-to-T_ratio}
\end{figure}

In Fig.~\ref{SF_ratios}(b), we display the longitudinal-to-transverse ratio 
of the structure functions $R_{L/T} = F_L (x,Q^2) / F_T (x,Q^2)$, 
where $F_T = F_2 - F_L$, for $Q^2 = 1$ and 10~GeV$^2$ in the two approaches.
It is seen that the four ratios slightly increase with $x$ and with $Q^2$.
These behaviors agree qualitatively with those observed in 
the previous analysis of the nucleon DIS~\cite{Watanabe:2013spa}.
Furthermore, we show in Fig.~\ref{L-to-T_ratio}
the $Q^2$ dependence of the ratio for a fixed $x=10^{-3}$.
The behaviors of the two curves are similar to each other, and 
both increase with $Q^2$. Although experimental data are not yet available
for comparison at this moment, our predictions for $F_L$ and for the scale dependence
indicated in Fig.~\ref{L-to-T_ratio} can be tested at linear colliders 
in the future.

%%%%%%%%%%%%%%%%%%%%%%%%%%%%%%%%%%%
\section{Summary}
%%%%%%%%%%%%%%%%%%%%%%%%%%%%%%%%%%%
In this work, we have investigated the electron-photon DIS at 
small $x$ in holographic QCD. In addition to the photon-photon scattering, 
we have also taken into account the photon-$\rho$ meson scattering based on 
the vector meson dominance model. The structure functions obtained in 
the two approaches are consistent with each other, and with the 
LEP data and those from the PDF parameterization of the 
real photon in the considered kinematic region.
The density distribution of the quasi-real photon,  
described by Eq.~\eqref{P13_2} with a tiny four-momentum squared, 
has a substantial component from the UV to IR 
region, while that of the $\rho$ meson has most of its component in 
the IR region. Hence, our findings indicate a nontrivial realization of 
the vector meson dominance in our model setup.
The fact that the present holographic formalism works well 
in both the studies of the photon and nucleon structures 
strongly supports its further applications to other high energy
scattering processes, to which the conventional
perturbation technique does not apply.

\bibliographystyle{JHEP}
\bibliography{hQCD}

%\begin{thebibliography}{99}
%\bibitem{...}
%....

%\end{thebibliography}

\end{document}